# Electron Emission in Superfluid and Low-temperature Vapor Phase Helium

## Isaac F. Silvera[1] and Jacques Tempere[1,2]

[1]Lyman Laboratory of Physics, Harvard University, Cambridge MA 02138, USA.

[2]TFVS, Universiteit Antwerpen, Groenenborgerlaan 171, 2020 Antwerpen, Belgium

Tungsten filaments used as sources of electrons in a low-temperature liquid or gaseous helium environment have remarkable properties of operating at thousands of degrees Kelvin in surroundings at temperatures of order 1 K. We provide an explanation of this performance in terms of important changes in the thermal transport mechanisms. The behavior can be cast as a first-order phase transition.

Electrons and ions in low-temperature helium have fascinated physicists for several decades [1, 2]. Phenomena such as two-dimensional electron gases on the surface of liquid helium, Wigner crystallization [3, 4] and quantum melting [5], single-electron bubbles [6, 7], multi-electron bubbles [8], electrical discharge and breakdown, etc. continue to reveal fascinating properties with unusual explanations. There are several sources of electrons and are not only useful, but the physics of these sources themselves is interesting. Some sources in a low temperature environment are nuclear beta-decay where a high-energy beta particle can create electrons by ionization of helium atoms, photoemission, field emission or Fowler-Nordheim emission, where electrons are emitted from a sharp conducting tip at a large negative potential, corona discharge [9], and thermionic emission from a tungsten wire. There are a number of unusual observations in helium such as electrical breakdown in superfluid helium and electron emission from filament wires. Although electrical breakdown in a gas can occur at a relatively low voltage, one might expect liquid helium to withstand extremely high voltages. Recently Sethumadhavan et al [10] showed that electrical breakdown between electrodes (one in the liquid and the other in the vapor phase) in low-temperature liquid helium (100 mK) occurred at substantially lower than anticipated voltages due to Penning ionization of helium excimers in surface states. The excimers were produced by free electrons in the helium, created by a beta source and accelerated by the electrode potentials. The tungsten filament can operate at temperatures of a few thousand Kelvin, glowing brilliantly due to blackbody radiation and emitting electrons by thermionic emission in a helium environment that remains at $T \sim 1K$ [11]. This source has been a "workhorse" for experiments on electrons, yet there has not been an elucidation of this phenomenon. The explanation involves a complex interplay of electrical properties of the filament and heat transport mechanisms in the helium. In this letter we explain this unusual behavior and show that it can be described as a first-order phase transition.

Tungsten filaments are well-known as electron sources in vacuum tubes. A current is passed through the filament wire suspended from two electrodes in a vacuum. As the current increases the filament warms due to ohmic heating and its temperature dependent resistance (R) increases. The two ends are thermally anchored at the electrodes that are at a modest temperature. The filament has a somewhat parabolic temperature profile as the ohmic heating in the wire is conducted from the hotter middle region to the electrodes, as well as being dissipated by blackbody radiation. Eventually the center becomes glowing hot at a few thousand K and the filament emits electrons according to the well-known Richardson law [12]. An I-V characteristic is shown by the dashed line in Fig. 1, where the slope is the resistance R. For this curve, with the filament operating in a vacuum, the peak central temperature increases continuously with increasing current. The behavior is very different in low temperature helium gas or liquid. Spangler and Hereford [11] showed that a tungsten filament with a diameter of several microns (~1 cm long) immersed in superfluid helium could be operated to ~2500 K with electron emission currents up to 0.5 µA. When sufficiently heated, a stable vapor sheath with a radius of ~100 microns [13] formed around the filament to insulate it from the superfluid helium, accompanied by an audible hissing sound. Figure 1 shows data we have taken for a filament operating in the vapor phase of a low temperature helium cell (T~1.3 K). Date et al [14] observed a negative resistance region such as in Fig. 1 for a filament in superfluid helium, but with a much sharper reduction in current (called the flyback) at or above the critical voltage $V_C$.

Most tungsten filament sources are operated in the vapor phase of helium above bulk superfluid helium, to produce electrons on the helium surfaces. To minimize the heating which can be several hunred millwatts (see Fig. 1) the usual mode of operation is to pulse the source on for a few seconds above the critical voltage. For the data in Fig. 1 a cell with high cooling power was used to enable continuous measurements of the steady state curve with a minimum warming of the helium. There are two branches of operation, the low-resistance and the high-resistance branches. If the filament is operated in a constant voltage mode (so that the current is free to vary according to the impedance of the circuit), when the voltage reaches a critical value, $V_c$, the current falls back to a lower value and the filament suddenly becomes hot and emits electrons, as we have measured with a low temperature collector. With increasing voltage the filament glows hotter and hotter with concomitant increasing electron emission. Reducing the voltage traces the curve back down to the cold, low-R branch. If the filament is operated in a constant current mode, then when $I_c$ is reached (see Fig. 1), the operating point jumps vertically to a higher value of voltage on the high-R branch, operating at a high current and emission, and is very hot. Summarizing the key features, the filament has two branches of operation: on the upper branch its distributed resistance is high, it is hot and emits electrons; on the lower branch its resistance is low, it is cool, and there is no measurable emission. In the constant voltage mode, at or above $V_c$ the resistance rapidly increases and since the voltage is constant, the current falls back. We now explain how changes in the mode of heat transport from the filament lead to this behavior.

Consider a segment [x, x+Δx] of the wire, with a temperature difference ΔT between the ends. The Fourier heat equation for ohmic heating along this segment can be written as

$$-A\kappa(x)\frac{\Delta T}{\Delta x} = i^2\frac{\rho(x)\Delta x}{A} - \varepsilon\sigma T^4(x)\pi D\Delta x - h_i\left[T(x) - T_\infty\right]\pi D\Delta x \qquad (1)$$

Here $A$ is the cross-sectional surface of the wire, $\rho(x)$ is the resistivity of the wire, and $\kappa(x)$ is its thermal conductivity. Both $\rho$ and $\kappa$ depend on the position along the wire through their temperature dependence. The first term on the right is due to ohmic heating. Besides ohmic heating, there are other heat flow processes (we ignore energy loss through thermionic emission itself.) The second term on the right represents radiative heat losses, where $\varepsilon$ and $\sigma$ are the emissivity and Stefan's constant, respectively, and $\pi D\Delta x$ is the surface area of the wire segment $\Delta x$ with diameter $D$. The third term represents losses to the environment through heat conduction into the gas or liquid with $T_\infty$ being the ambient temperature of the (cryogenic) environment, and $h_i$ a constant representing transport of heat from the wire into the surrounding helium environment. For the present purposes we set $T_\infty$ to 0 K.

The key difference in the behavior of the filament in a vacuum or in a cold gas or liquid environment is the thermal transport of energy from the wire to the gas (liquid). As the filament heats, the heat transport mechanism changes from a high thermal conductivity convective mode to a low conductivity chaotic diffusive mode. We model our system as two concentric surfaces of diameters $D$ and $D_\infty$, with the latter being $D_\infty \sim 1$ cm, where the temperature has relaxed to $T_\infty$. Then for diffusive heat flow [15]

$$\dot{Q} = \frac{2\pi\Delta x\Delta T}{\ln(D_\infty / D)}\kappa_{He},$$

or

$$h_d = \frac{2\kappa_{He}}{D\ln(D_\infty / D)}. \qquad (2)$$

where $\kappa_{He}$ is the thermal conductivity of helium gas. We include both the diffusive (conductive) and convective heat transport with the Nusselt number Nu, the ratio of the convective to the conductive heat transport, so

$$h_i = \frac{2\kappa_{He}}{D\ln(D_\infty / D)} Nu_i, \qquad\qquad (3)$$

with $Nu_i = 1$ for the diffusive heat transport and we set $Nu_i = 4$ for the convective heat transport [16]. The exact value of this ratio is not too important as in our calculation we want to show that there is a change of resistance of the tungsten filament with a change in the heat transfer mechanism. The resulting differential equation that we solve numerically, using Mathematica, is

$$\frac{d}{dx}\left[\kappa(x)\frac{dT(x)}{dx}\right] = i^2 \frac{\rho(x)}{\left(\pi D^2 / 4\right)^2} - \varepsilon\sigma\pi T^4(x)\frac{4}{D} - h_i T(x)\frac{4}{D} \quad . \qquad (4)$$

The resistivity of tungsten increases almost linearly with temperature at the relevant temperatures and is fit to an appropriate function; the thermal conductivity of the tungsten wire is taken constant. The resulting temperature distributions in the limit of small heat transport to the surrounding helium environment are consistent with those obtained by Durakiewicz and Halas [17] for filaments heated in vacuum. The equation can be solved for the low-R and high-R branches for a filament in helium (see ahead). An almost Gaussian temperature profile, shown in Fig. 2, is found for the hot branch.

The solutions to Eq. 4 depend sensitively on the thermal conductivity of the environment, which is important for heat removal from the filament. Although a solution could be found for some conditions, the numerical procedure did not always easily converge. The calculation of the temperature at the bending point of the temperature profile converges easily, and the obtained dependence of this temperature on the system parameters agrees well with the full solution in the cases where the differential equation was solvable. This allows the use of the temperature at the bending point as a control parameter in the full calculations. In Fig. 2 we show the peak temperature of the filament as a function of filament current for three values of the helium-filament thermal conductivity, $\kappa_{He}$. We see that depending on the value of $\kappa_{He}$ and the current, the peak temperature can be low, of the order of the helium environment, or as high as a few thousand Kelvin. The thermal transport to the environment plays an important role for the heat loss; at $T = 1500$ K it is larger than the radiative losses. Thus, if operating at a given current, say 150 mA on the central curve in Fig. 2 and the thermal transport is reduced, the system jumps up to the lower heat transport curve at a much higher temperature. The equations could also be solved to show the abrupt changes in current or voltage at the critical values shown in Fig. 1; this is shown in Fig. 3. The calculated power dissipated by the filament when hot compared favorably with experimental observations.

It is now easy to understand the behavior of the filament immersed in superfluid liquid helium. At low temperature the filament is in intimate contact with the superfluid, the best known thermal conductor, which efficiently removes heat. At a certain value of current the liquid can no longer carry the heat and the vapor sheath, with a much smaller thermal conductivity, abruptly develops around the wire. The filament now operates at a different point in Fig. 2, with lower thermal conductivity, so that the peak temperature jumps to a high value. In this case of immersion of the wire in superfluid helium, the change in the Nusselt number is probably larger than in our assumption of 4 going to 1. Now consider the operation of the filament in the vapor phase at low temperature. At very low filament temperature the heat removal is by convective cooling due to the superfluid helium film or the vapor as the helium film is burnt off of the filament, and goes over to convective cooling in the dense surrounding gas. As the temperature of the filament increases the flow morphology of the helium gas changes to chaotic behavior and the Nusselt number (characterizing the thermal conductivity) drops [18]. Again, the behavior is similar to the filament immersed in superfluid helium, and the temperature rises abruptly, accompanied by electron emission.

A phase transition, for example, the normal to superfluid helium transition is characterized by an order parameter and an order variable. The order parameter is the important observable, the normalized superfluid density, and the variable is the temperature. In a first order phase transition the free energy has two branches corresponding to the normal and superfluid phases. We can generalize this concept to characterize the behavior of the filament. The two resistance branches of the filament are similar to a two-phase system, and we can describe the behavior of the filament as a first-order phase transition. Clearly the shift of the operating point from one branch to another is due to changes in the properties of the helium environment; in the case of the liquid a vapor sheath is formed and in the vapor phase there is a change in the flow morphology. To describe this as a phase transition for the two cases we take as the order parameter the important observable, the electron emission current normalized to its maximum value. This does not imply that there is a phase transition in the wire; rather there is a change in the environment. This change leads to an increase in the thermionic emission of the wire, so that the emission current can be used as a parameter characterizing the change in the environment. The order variable is the thermal transport coefficient, $h$. The change in $h$ can be quite abrupt, as for example the change that occurs with the sudden development of a vapor sheath for a filament immersed in superfluid helium. Thus, one expects a first order type of phase transition, linked to the transition to a sheath, or in convecting helium flow from harmonic to chaotic heat transport in the vapor phase, studied by Castillo and Hoover [19]. The ordering in the filament emission only takes place when the filament's critical voltage or current are surpassed.

We thank the DOE Grant No. DE-FG02-ER45978 and the FWO-V Projects Nos. G.0435.03, G.0306.00, G.0274.01N for support of this research. Jonathan Sommer and Jian Huang assisted with some of the data taking.

**Figure Captions**

Figure 1.  The experimental I-V characteristic of a 12.5 micron diameter tungsten filament operating at low temperature in the vapor phase of helium.   The dense set of data points is for constant voltage operation, and is a steady state set of points for slow variation of the voltage.  The vertical dotted line is the path for a constant current mode (data not shown).  This data was for a thoriated filament with a lower work function than pure tungsten.  The dashed curve is the path that would be followed for a filament operating in vacuum.  $V_C$ and $I_C$ are the critical voltage and current for a dramatic behavioral change where the filament becomes hot and starts emitting electrons.

Figure 2.  The calculated temperature of a 25 micron diameter tungsten wire in the vicinity of the peak temperature as a function of current through the wire.  The different curves are for different heat transfer coefficients between the wire and the helium, using values four times larger and four times smaller than a selected standard value.  The inset shows the calculated temperature profile of the wire for currents of 140 to 220 mA, in steps of 20 mA.  Note the sharpening of the profile as the peak temperature increases.

Figure 3.  The calculated voltage-current characteristic as the current or voltage is kept fixed and the Nusselt number transitions from high to low value, shown in the inset.  In this case it was varied by a factor of 3.  The main characteristics of the experimental curve (see Fig. 1) are nicely reproduced in this model.  This simulation is closer to what we expect for the filament operating in the vapor phase.  For operation in the liquid phase of helium a much sharper transition (current flyback) is observed when the vapor sheath is formed.

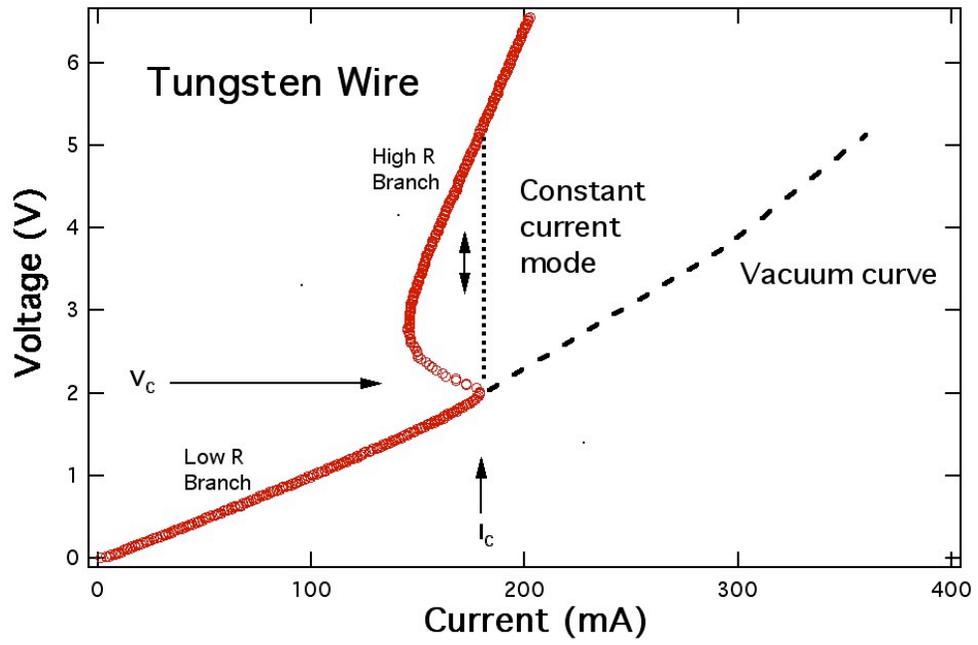

Figure 1, Silvera PRL

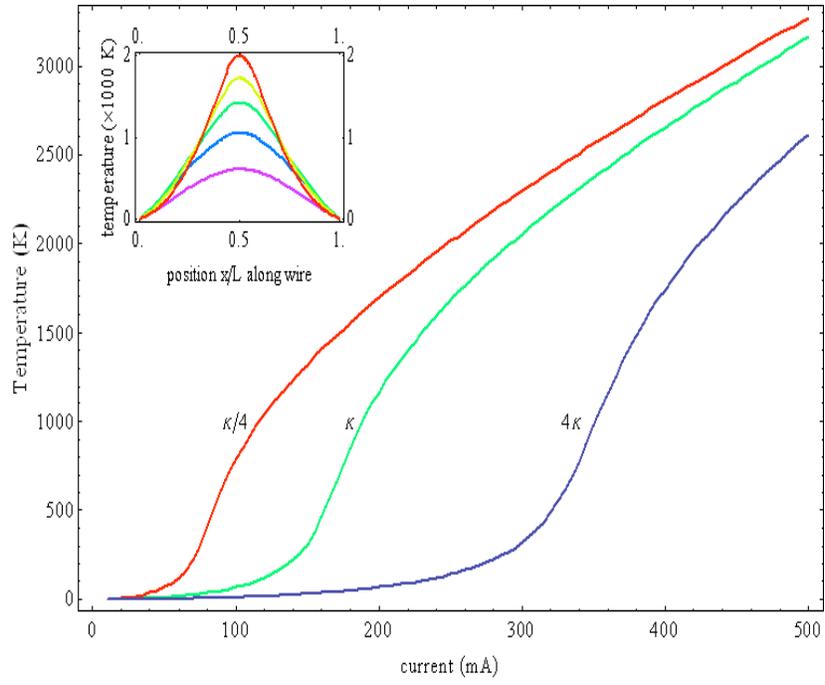

Figure 2, Silvera, PRL

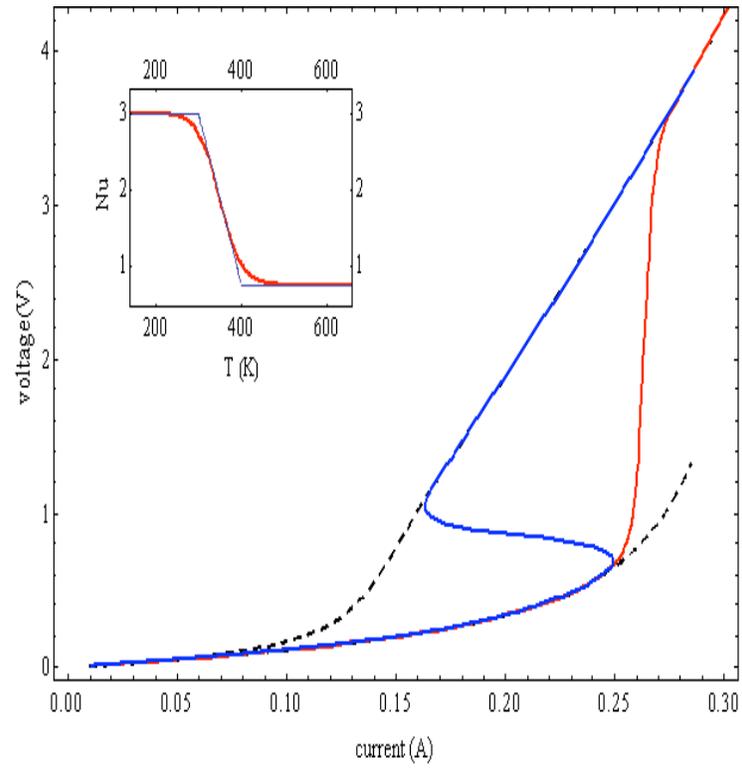

Figure 3, Silvera, PRL